\documentclass{article}
\usepackage{spconf,amsmath,graphicx,amsfonts,multirow,booktabs,float,threeparttable,lipsum}

\newcommand{\tabincell}[2]{\begin{tabular}{@{}#1@{}}#2\end{tabular}}

\title{Within-sample Variability-invariant loss for robust speaker recognition under noisy environments}
\name{Danwei Cai$^{\star}$, Weicheng Cai$^{\dagger}$, Ming Li$^{\star \dagger}$}
\address{$^{\star}$Department of Electrical and Computer Engineering, Duke University, Durham, USA\\$^{\dagger}$Data Science Research Center, Duke Kunshan University, Kunshan, China \\\texttt{ming.li369@duke.edu}}

\begin{document}
\ninept

\maketitle

\begin{abstract}
Despite the significant improvements in speaker recognition enabled by deep neural networks, unsatisfactory performance persists under noisy environments. In this paper, we train the speaker embedding network to learn the ``clean'' embedding of the noisy utterance. Specifically, the network is trained with the original speaker identification loss with an auxiliary within-sample variability-invariant loss. This auxiliary variability-invariant loss is used to learn the same embedding among the clean utterance and its noisy copies and prevents the network from encoding the undesired noises or variabilities into the speaker representation. Furthermore, we investigate the data preparation strategy for generating clean and noisy utterance pairs on-the-fly. The strategy generates different noisy copies for the same clean utterance at each training step, helping the speaker embedding network generalize better under noisy environments. Experiments on VoxCeleb1 indicate that the proposed training framework improves the performance of the speaker verification system in both clean and noisy conditions.
\end{abstract}

\section{Introduction}
Automatic speaker verification (ASV) refers to automatically making the decision to accept or reject a claimed speaker by analyzing the given speech from that speaker. In the past few years, the performance of ASV systems has been improved significantly with the successful application of deep neural network (DNN) to speaker embedding modeling \cite{snyder_x-vectors:_2018, cai_exploring_2018}. However, unsatisfactory performance persists under noisy environments, which commonly noticed in smartphones or smart speakers with ASV applications. The additive noises on a clean speech contaminate the low energy regions of the spectrogram and blur the acoustic details \cite{woelfel_distant_2009}. These noises result in the loss of speech intelligibility and quality, imposing great challenges on speaker recognition systems.

To compensate for these adverse impacts, various approaches have been proposed at different stages of the ASV systems. At the signal level, DNN based speech or feature enhancement \cite{zhao_robust_2014, kolboek_speech_2016, oo_dnn-based_2016, plchot_audio_2016} has been investigated for ASV under complex environment. At the feature level, feature normalization techniques \cite{pelecanos_feature_2001} and noise-robust features such as power-normalized cepstral coefficients (PNCC) \cite{PNCC} have also been applied to ASV systems. At the model level, robust back-end modeling methods such as multi-condition training of probabilistic linear discriminant analysis (PLDA) models~\cite{garcia-romero_multicondition_2012} and mixture of PLDA~\cite{mak_mixture_2016} were employed in the i-vector \cite{dehak_front-end_2011} framework. Also, score normalization \cite{peer_reverberation_2008} could be used to improve the robustness of the ASV system under noisy scenarios.

More recently, researchers are working on training deep speaker networks to cope with the distortions caused by noise. Within this framework, there are two main methods. The first one regards the noisy data as a different domain from the clean data and applies adversarial training to deal with domain mismatch and get a noise-invariant speaker embedding \cite{zhou_training_2019, meng_adversarial_2019}. The second method employs a DNN speech enhancement network for ASV tasks. Shon \textit{et al.}~\cite{shon_voiceid_2019} train the speech enhancement network with feedbacks from the speaker network to find the time-frequency bins that are beneficial to ASV tasks with noisy speech. Zhao \textit{et al.}~\cite{zhao_robust_2019} uses the intermediate result of the speech enhancement network as an auxiliary input for the speaker embedding network and jointly optimize these two networks.

In this work, our network learns enhancement directly at the embedding level for speaker recognition under noisy environments. We train the deep speaker embedding network by incorporating the original speaker identification loss with an auxiliary within-sample loss. The speaker identification loss learns the speaker representation using the speaker label, while the within-sample loss aims to learn the embedding of noisy utterance as similar as possible to its clean version. In this way, the deep speaker embedding network is trained to prevent from encoding the additive noises into the speaker representation and learn the ``clean'' embedding for the noisy speech utterance. The loss that helps the speaker network to learn variability-invariant embedding is called within-sample variability-invariant loss. The similar idea has been applied in speech recognition \cite{liang_learning_2018} and far-field speaker recognition with reverberation \cite{intel_voices}.

Furthermore, to fully explore the modeling ability of the within-sample variability-invariant loss, we dynamically generate the clean and noisy utterance pairs when preparing data for the training process. Different noisy copies for the same clean utterance are generated at different training steps, helping the speaker embedding network generalize better under noisy environments. 


\section{Revisit: Deep speaker embedding}

In this section, we describe the deep speaker embedding framework, which consists of a frame-level local pattern extractor, an utterance-level encoding layer, and several fully-connected layers for speaker embedding extraction and speaker classification.

Given a variable-length input feature sequence, the local pattern extractor, which is typically a convolutional neural network (CNN)~\cite{cai_exploring_2018} or a time-delayed neural network (TDNN) \cite{snyder_x-vectors:_2018}, learns the frame-level representations. An encoding layer is then applied to the top of it to get the utterance level representation. The most common encoding method is the average pooling layer, which aggregates the statistics (i.e., mean, or mean and standard deviation)~\cite{snyder_x-vectors:_2018, cai_exploring_2018}. Self-attentive pooling layer~\cite{bhattacharya_deep_2017}, learnable dictionary encoding layer~\cite{cai_novel_2018}, and dictionary-based NetVLAD layer \cite{chen_end--end_2018, xie_utterance-level_2019} are other commonly used encoding layers. Once the utterance-level representation is extracted, a fully connected layer and a speaker classifier are employed to further abstract the speaker representation and classify the training speakers. After training, deep speaker embedding is extracted after the penultimate layer of the network for the given variable-length utterance.

In this work, the local pattern extractor is a residual convolutional neural network (ResNet) \cite{He2016Deep}, and the encoding layer is a global statistics pooling (GSP) layer. For the frame-level representation $ \mathbf{F} \in \mathbb{R}^{C\times H\times W}$, the output of GSP is a utterance-level representation $\mathbf {V}={\left[\mu_1, \mu_2, \cdots, \mu_C, \sigma_1, \sigma_2, \cdots, \sigma_C\right]}$, where $\mu_c$ and $\sigma_c$ are the mean and standard deviation of the $c^{\textrm{th}}$ feature map:
\begin{equation} \begin{aligned}
\mu_c = & \frac{1}{H\times W}\sum_{i=1}^{i=H}\sum_{j=1}^{j=W} \mathbf{F}_{c,i,j} \\
\sigma_c = & \sqrt{ \frac{1}{H\times W}\sum_{i=1}^{i=H}\sum_{j=1}^{j=W} (\mathbf{F}_{c,i,j} - \mu_c)^2 }
\end{aligned} \end{equation}
and $C, H, W$ denote the number of channels, height and width of the feature map respectively.

\section{Methods}
In this section, we describe the proposed framework with within-sample variability-invariant loss and online noisy data generation. 

\subsection{Within sample variability-invariant loss}

A clean speech and its noisy copies contain the same acoustic contents for recognizing speakers. Ideally, the speaker embeddings of the noisy utterance should be the same as its clean version. But in reality, the deep speaker embedding network usually encodes the noises as parts of the speaker representation for the noisy speech.

In this work, we train the local pattern extractor to learn the enhancement at the embedding level. Formally, for a clean utterance $\mathbf{s}_c$ and its noisy copy $\mathbf{s}_n = \mathbf{s}_c + \mathbf{n}$ with noise $\mathbf{n}$, the speaker embeddings $\mathbf{f}_c, \mathbf{f}_n \in \mathbb{R}^p$ extracted by the network $\mathbf{N}$ are
\begin{equation} \begin{aligned}
&\mathbf{f}_c = \mathbf{N}(\mathbf{s}_c) \\
&\mathbf{f}_n = \mathbf{N}(\mathbf{s}_n) = \mathbf{N}(\mathbf{s}_c + \mathbf{n})
\end{aligned}\end{equation}

A loss function $l$ on the embedding level is used to measure the difference between the noisy embedding and the clean embedding form the same sample. The learning objective for the speaker network is 
\begin{equation}
    \min l(\mathbf{f}_c,\mathbf{f}_n) = \min_{\mathbf{N}} l[\mathbf{N}(\mathbf{s}_c + \mathbf{n}), \mathbf{N}(\mathbf{s}_c)]
\end{equation} 
In this way, the speaker embedding network is trained to ignore the additive noises and learn noise-invariant embeddings. We refer this loss function as within-sample variability-invariant loss. Two different loss functions are investigated in this work, i.e., mean square error (MSE) regression loss and cosine embedding loss.

The MSE regression loss calculates the mean of the square \textit{L}2 norm between the clean embedding $\mathbf{f}_c$ and its noisy version $\mathbf{f}_n$,
\begin{equation}\begin{aligned}
    l_\textrm{MSE} = & \frac{1}{p} \parallel\mathbf{f}_c-\mathbf{f}_n\parallel_2^2 \\
    = & \frac{1}{p} \parallel \mathbf{N}(\mathbf{s}_c + \mathbf{n}) - \mathbf{N}(\mathbf{s}_c)\parallel_2^2 
\end{aligned}\end{equation} 
where $\parallel\cdot\parallel_2$ denotes the \textit{L}2 norm, $p$ is the dimension of the speaker embeddings $\mathbf{f}_c, \mathbf{f}_n$.

The cosine embedding loss calculates the cosine distance between the clean embedding $\mathbf{f}_c$ and its noisy version $\mathbf{f}_n$,
\begin{equation}\begin{aligned}
    l_\textrm{cos} = & 1-\cos (\mathbf{f}_c,\mathbf{f}_n)\\
    = & 1-\cos[\mathbf{N}(\mathbf{s}_c + \mathbf{n}), \mathbf{N}(\mathbf{s}_c)]
\end{aligned}\end{equation}

\begin{figure}[t]
  \centering
  \includegraphics[width=0.793\linewidth]{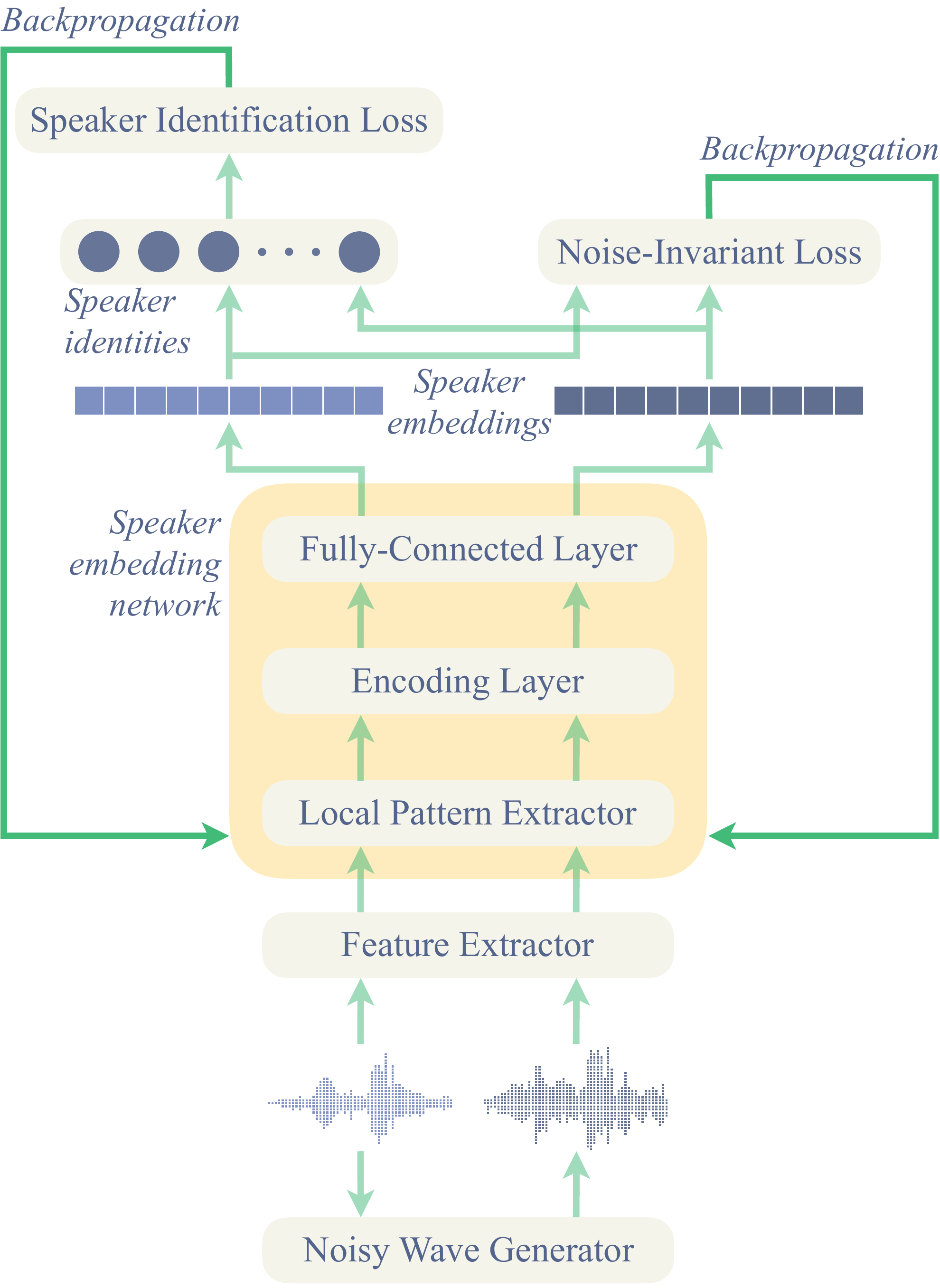}
  \caption{Training deep speaker embedding network with within-sample variability-invariant loss.}
  \label{fig: e2e}
\end{figure}

The within-sample variability-invariant loss works with the original speaker identification loss together to train the speaker embedding network. The speaker identification loss is typically a cross-entropy. In our implementation, the hyper-parameters of the network are updated twice at each training step. The first update from the speaker identification loss is followed by the second update from the within-sample variability-invariant loss. Figure \ref{fig: e2e} shows the flowchart of our proposed framework.

\begin{table*}[t]
    \caption{Performance on Voxceleb1 test set (DCF and EER[\%]), s. denotes softmax, As. denotes A-softmax. The bold highlight the best DCF and EER for the speaker networks trained with softmax and A-softmax.}
    \label{tab: results}
    \centering
    \begin{threeparttable}
    \begin{tabular}{@{\ \ }c@{}c c@{\ \ }r |c@{\ \ }r |c@{\ \ }r |c@{\ \ }r |c@{\ \ }r  |c@{\ \ }r |c@{\ \ }r |c@{\ \ }r@{\ \ }}
    \toprule 
    \multirow{2}*{\tabincell{c}{\textbf{Noise} \\ \textbf{Type}}} & \multirow{2}*{\textbf{SNR}} & \multicolumn{2}{c}{\makebox[0.1mm][c]{\textbf{Clean}}} & \multicolumn{2}{c}{\makebox[0.1mm][c]{\textbf{Offline AUG}}} & \multicolumn{2}{c}{\makebox[0.1mm][c]{\textbf{Online AUG}}} & \multicolumn{2}{c}{\makebox[0.1mm][c]{\textbf{Online AUG}}} & \multicolumn{2}{c}{\makebox[0.1mm][c]{\textbf{Online AUG}}} & \multicolumn{2}{c}{\makebox[0.1mm][c]{\textbf{Online AUG}}} & \multicolumn{2}{c}{\makebox[0.1mm][c]{\textbf{Online AUG}}} & \multicolumn{2}{c}{\makebox[0.1mm][c]{\textbf{Online AUG}}}\cr

   
   & & \multicolumn{2}{c}{\makebox[0.1mm][c]{\textbf{softmax}}} & \multicolumn{2}{c}{\makebox[0.1mm][c]{\textbf{softmax}}} & \multicolumn{2}{c}{\makebox[0.1mm][c]{\textbf{softmax}}} & \multicolumn{2}{c}{\makebox[0.1mm][c]{\textbf{s.+MSE}}} & \multicolumn{2}{c}{\makebox[0.1mm][c]{\textbf{s.+cosine}}} & \multicolumn{2}{c}{\makebox[0.1mm][c]{\textbf{A-softmax}}} & \multicolumn{2}{c}{\makebox[0.1mm][c]{\textbf{As.+MSE}}} & \multicolumn{2}{c}{\makebox[0.1mm][c]{\textbf{As.+cosine}}} \cr
    \midrule
    \multicolumn{2}{c}{Original set} & 0.453 & 3.73 & 0.451 & 3.65 & 0.516 & 3.66 & \textbf{0.418} & \textbf{3.46} & 0.459 & 3.47 & 0.456 & 3.56 & 0.442 & 3.49 & \textbf{0.435} & \textbf{3.12} \\

    \midrule
    \multirow{5}*{Babble} 
&  0 & 0.974 & 24.16 & 0.900 & 13.29 & 0.877 & 12.32 & 0.822 & \textbf{11.10} & \textbf{0.821} & 11.21 & 0.861 & 12.57 & \textbf{0.844} & \textbf{10.93} & 0.848 & 11.78 \\
&  5 & 0.881 & 12.25 & 0.749 & 6.96 & 0.688 & 6.63 & \textbf{0.683} & \textbf{5.94} & 0.709 & 5.99 & 0.647 & 6.56 & 0.662 & \textbf{5.83} & \textbf{0.619} & 5.97 \\
& 10 & 0.682 & 6.91 & 0.588 & 5.23 & 0.577 & 4.87 & \textbf{0.535} & \textbf{4.57} & 0.548 & 4.68 & \textbf{0.519} & 4.86 & 0.610 & \textbf{4.38} & 0.557 & 4.44 \\
& 15 & 0.596 & 4.94 & 0.506 & 4.46 & 0.538 & 4.27 & 0.508 & \textbf{3.94} & \textbf{0.479} & 4.13 & \textbf{0.476} & 4.15 & 0.509 & 3.89 & 0.480 & \textbf{3.73} \\
& 20 & 0.493 & 4.07 & 0.483 & 4.05 & 0.513 & 3.76 & \textbf{0.440} & \textbf{3.61} & 0.484 & 3.75 & 0.467 & 3.77 & 0.478 & 3.66 & \textbf{0.453} & \textbf{3.36} \\

    \midrule
    \multirow{5}*{Music}
&  0 & 0.921 & 16.02 & 0.758 & 9.01 & 0.728 & 8.44 & \textbf{0.710} & \textbf{7.65} & 0.742 & 7.74 & 0.784 & 8.66 & 0.725 & \textbf{7.27} & \textbf{0.722} & 7.79 \\
&  5 & 0.838 & 9.81 & 0.665 & 6.02 & 0.678 & 5.92 & 0.608 & 5.47 & \textbf{0.582} & \textbf{5.29} & 0.628 & 5.88 & \textbf{0.594} & 5.36 & 0.626 & \textbf{5.23} \\
& 10 & 0.691 & 6.31 & 0.560 & 4.90 & 0.577 & 4.67 & 0.572 & \textbf{4.30} & \textbf{0.542} & 4.51 & 0.510 & 4.56 & 0.507 & 4.25 & \textbf{0.490} & \textbf{4.11} \\
& 15 & 0.547 & 4.82 & 0.508 & 4.29 & 0.519 & 4.15 & \textbf{0.458} & \textbf{3.90} & 0.476 & 3.94 & 0.484 & 4.05 & 0.479 & 3.82 & \textbf{0.456} & \textbf{3.63} \\
& 20 & 0.535 & 4.19 & 0.491 & 3.91 & 0.507 & 3.84 & \textbf{0.451} & 3.71 & 0.483 & \textbf{3.66} & 0.470 & 3.74 & 0.448 & 3.65 & \textbf{0.437} & \textbf{3.30} \\

    \midrule
    \multirow{5}*{Noise}
&  0 & 0.968 & 15.20 & 0.781 & 8.61 & 0.757 & 8.09 & 0.715 & \textbf{7.25} & \textbf{0.708} & 7.31 & \textbf{0.696} & 8.00 & 0.724 & \textbf{7.31} & 0.742 & 7.34 \\
&  5 & 0.823 & 9.81 & 0.675 & 6.43 & 0.688 & 6.03 & \textbf{0.629} & \textbf{5.56} & 0.637 & 5.62 & 0.657 & 6.09 & \textbf{0.615} & \textbf{5.64} & 0.640 & 5.65 \\
& 10 & 0.724 & 7.15 & 0.598 & 5.07 & 0.602 & 4.92 & \textbf{0.557} & 4.52 & 0.570 & \textbf{4.50} & 0.563 & 4.85 & 0.574 & 4.59 &\textbf{0.553} & \textbf{4.35} \\
& 15 & 0.611 & 5.54 & 0.556 & 4.50 & 0.579 & 4.38 & \textbf{0.492} & \textbf{4.11} & 0.521 & 4.14 & 0.519 & 4.30 & 0.528 & 4.03 & \textbf{0.503} & \textbf{3.85} \\
& 20 & 0.540 & 4.57 & 0.500 & 4.07 & 0.547 & 3.97 & \textbf{0.476} & 3.83 & 0.501 & \textbf{3.79} & 0.467 & 3.85 & 0.470 & 3.72 & \textbf{0.452} & \textbf{3.44} \\

    \midrule
    \multicolumn{2}{c}{All noises}
     & 0.798 & 9.40 & 0.644 & 6.33 & 0.650 & 6.00 & \textbf{0.602} & \textbf{5.51} & 0.614 & 5.56 & 0.607 & 6.01 & 0.607 & \textbf{5.40} & \textbf{0.596} & 5.45 \\
    \bottomrule
    \end{tabular}
    \end{threeparttable}
\end{table*}

\subsection{Online data augmentation}
In this work, we implement an online data augmentation strategy. Different parameters of noise types, noise clips and signal-to-noise ratio (SNR) are randomly selected to generate the clean-noisy utterance pair when training. Different permutations of these random parameters generate different noisy segments for the same utterance at different training steps, so the network never ``sees'' the same noisy segment from the same clean speech. 

During training, the SNR is a continuous random variable uniformly distributed between 0 and 20dB, and there are four types of noise: music, ambient noise, television, and babble. The television noise is generated with one music file and one speech file. The babble noise is constructed by mixing three to six speech files into one, which results in overlapping voices simultaneously with the foreground speech.
\begin{table}[t]
    \footnotesize
    \caption{\footnotesize The network architecture, $\mathbf{C}$(kernal size, stride) denotes the convolutional layer, $\mathbf{S}$(kernal size, stride) denotes the shortcut convolutional layer, $\left[\cdot \right]$ denotes the residual block.}
    \centering
    \begin{tabular}[c]{lcl}
        \toprule
        \textbf{Layer} & \textbf{Output Size} & \textbf{Structure} \\
        \midrule
        Conv1 & $16 \times 64 \times L$ & $\mathbf{C}(3\times 3, 1)$ \\
        \midrule
        \tabincell{l}{Residual\\Layer 1} & $16 \times 64 \times L$ & $\begin{bmatrix}
            \mathbf{C}(3\times 3, 1) \\
            \mathbf{C}(3\times 3, 1)
        \end{bmatrix} \times 3$ \\
        \midrule
        \tabincell{l}{Residual\\Layer 2} & $32 \times 32 \times \frac{L}{2}$ & $\begin{bmatrix}
            \mathbf{C}(3\times 3, 2) \\
            \mathbf{C}(3\times 3, 1) \\
            \mathbf{S}(1\times 1, 2)
        \end{bmatrix} \begin{bmatrix}
            \mathbf{C}(3\times 3, 1) \\
            \mathbf{C}(3\times 3, 1)
        \end{bmatrix}\times 3$ \\
        \midrule
        \tabincell{l}{Residual\\Layer 3} & $64 \times 16 \times \frac{L}{4}$ & $\begin{bmatrix}
            \mathbf{C}(3\times 3, 2) \\
            \mathbf{C}(3\times 3, 1) \\
            \mathbf{S}(1\times 1, 2)
        \end{bmatrix} \begin{bmatrix}
            \mathbf{C}(3\times 3, 1) \\
            \mathbf{C}(3\times 3, 1)
        \end{bmatrix}\times 5$ \\
        \midrule
        \tabincell{l}{Residual\\Layer 4} & $128 \times 8 \times \frac{L}{8}$ & $\begin{bmatrix}
            \mathbf{C}(3\times 3, 2) \\
            \mathbf{C}(3\times 3, 1) \\
            \mathbf{S}(1\times 1, 2)
        \end{bmatrix} \begin{bmatrix}
            \mathbf{C}(3\times 3, 1) \\
            \mathbf{C}(3\times 3, 1)
        \end{bmatrix}\times 2$ \\
        \midrule
        Encoding & $256$ & Global Statistics Pooling \\
        \midrule
        Embedding & $128$ & Fully Connected Layer\\
        Classifier & $1211$ & Fully Connected Layer\\
        \bottomrule
    \end{tabular}
    \label{table: architecture}
\end{table}

\section{Experiments}
\subsection{Dataset}
The experiments are conducted on Voxceleb 1 dataset \cite{nagrani_voxceleb:_2017}. The training data contain 148642 utterances from 1211 speakers. In the test data, 4874 utterances from 40 speakers construct 37720 test trials. Although the Voxceleb dataset collected from online video is not strictly in clean condition, we assume the original data as a clean dataset and generate noisy data from the original data.

The MUSAN dataset \cite{musan} is used as the noise source. We split the MUSAN into two non-overlapping subsets for training and testing noisy data generation respectively.

\subsection{Experimental setup}
Speech signals are firstly converted to 64-dimensional log Mel-filterbank energies and then fed into the speaker embedding network. The detailed network architecture is in table \ref{table: architecture}. The front-end local pattern extractor is based on the well known ResNet-34 architecture~\cite{He2016Deep}. ReLU activation and batch normalization are applied to each convolutional layer.

For the speaker identification loss, a standard softmax-based cross-entropy loss or angular softmax loss (A-softmax) \cite{liu_sphereface:_2017} is used. When training with softmax loss, dropout is added to the penultimate fully-connected layer to prevent overfitting. 

Three training data settings are investigated: (1) original Voxceleb~1 dataset (clean); (2) original training dataset and offline generated noisy data, i.e., the noisy data are generated in advance (offline AUG); (3) original training data with online data augmentation (online AUG).

At the testing stage, cosine similarity is used for scoring. We use equal error rate (EER) and detection cost function (DCF) as the performance metric. The reported DCF is the average of two minimum DCFs when $P_{\textrm{target}}$ is 0.01 and 0.001.

\subsection{Experimental results}
Eight deep speaker embedding networks are trained based on three training conditions and different loss functions. Table \ref{tab: results} shows the DCF and EER of three noise types (babble, ambient noise and music) at five SNR settings (0, 5, 10, 15, 20dB). Also, all of the 15 noisy testing trials are combined to form the ``all noises'' trial.

\begin{figure}[t!]
\centering
  \includegraphics[width=0.85\linewidth]{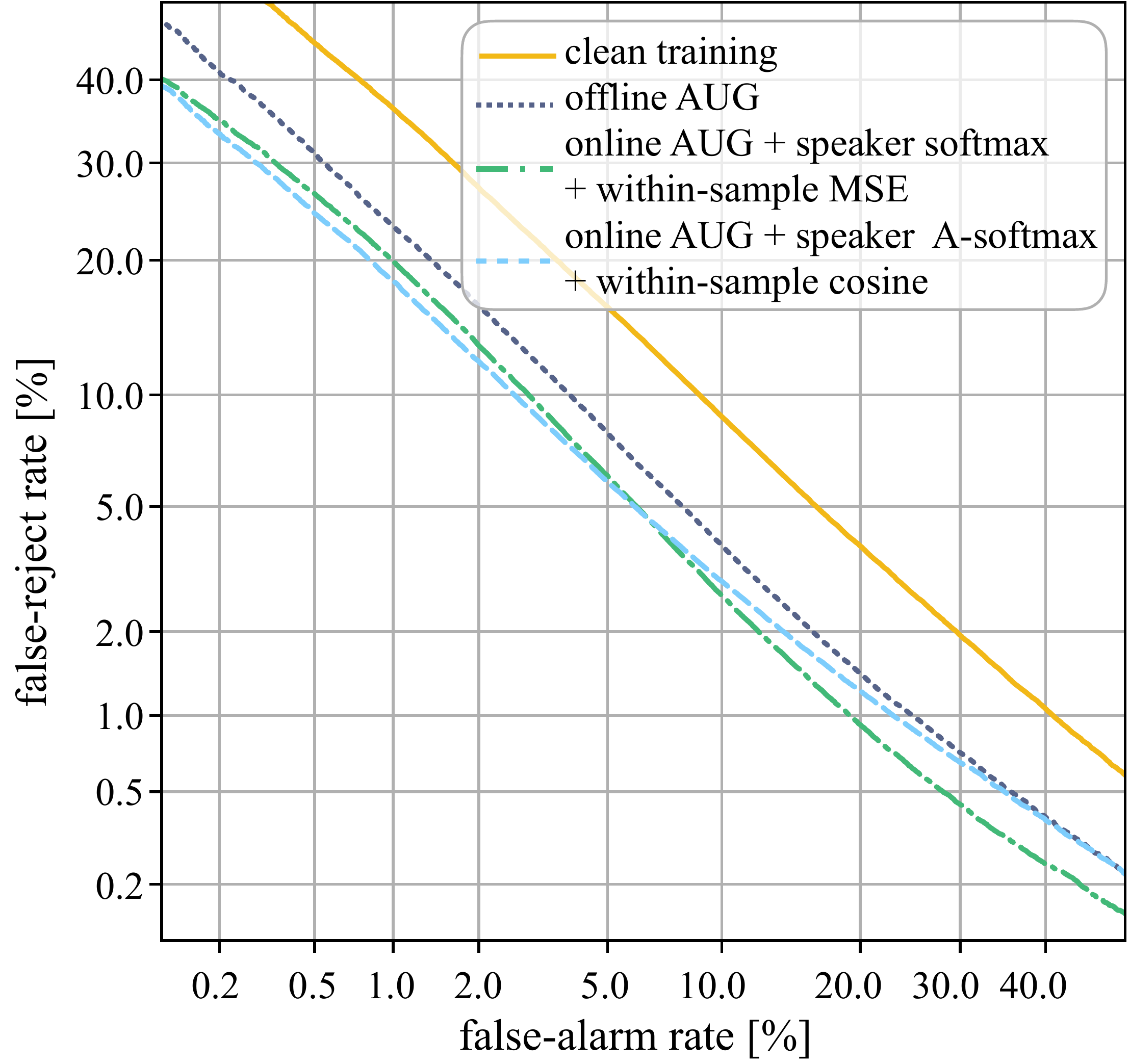}
  \caption{DET curves for four deep speaker embedding systems.}
  \label{fig:det_curve}
  \vspace*{\floatsep}

  \includegraphics[width=0.85\linewidth]{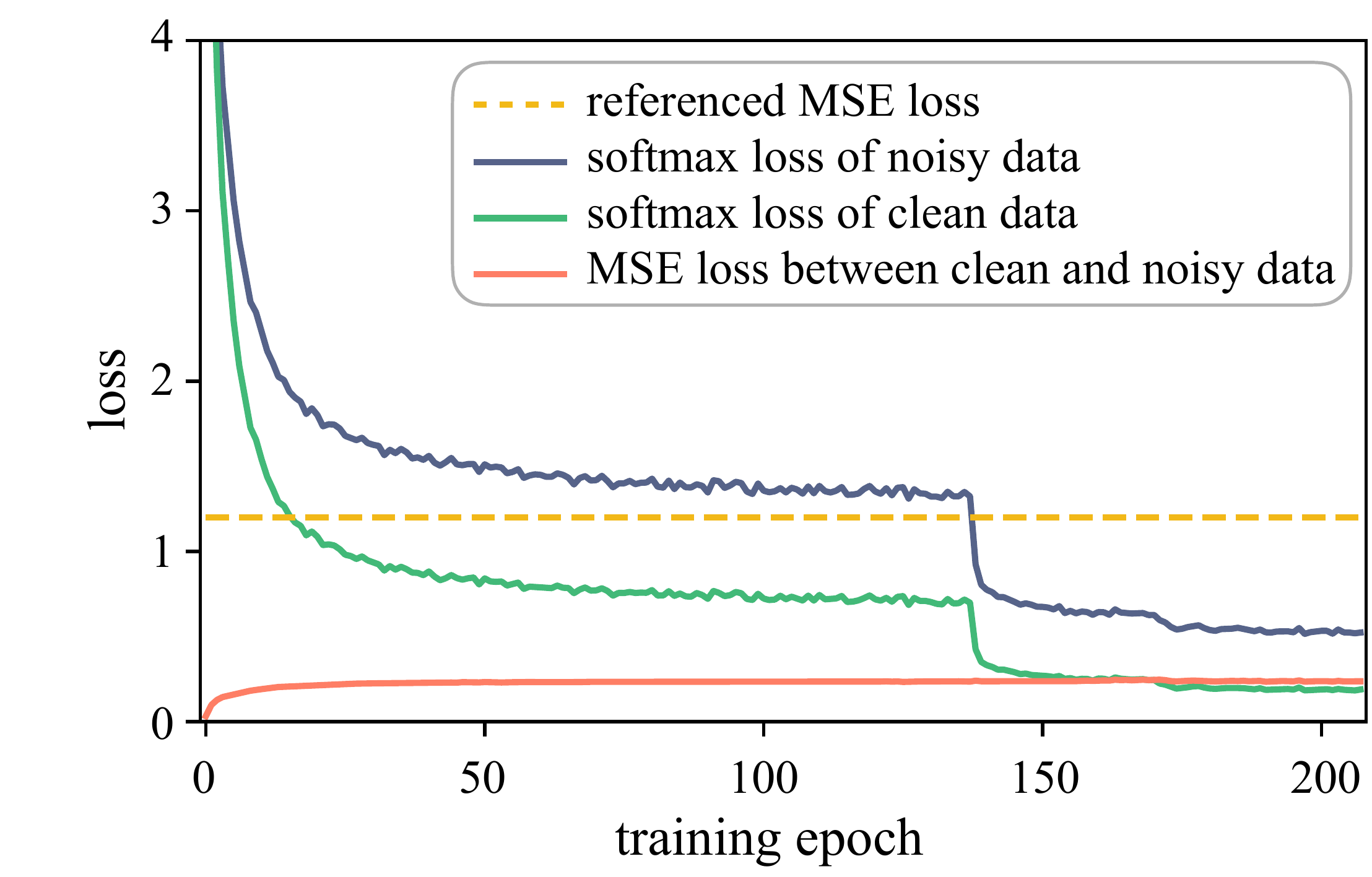}
  \caption{Three training loss curves for the network trained with speaker softmax loss and within-sample MSE loss. The referenced within-sample MSE loss between the clean and noisy data of the converged network trained with only softmax loss is also given.}
  \label{fig:loss}
\end{figure}

\begin{figure}[t]
\centering
  \includegraphics[width=\linewidth]{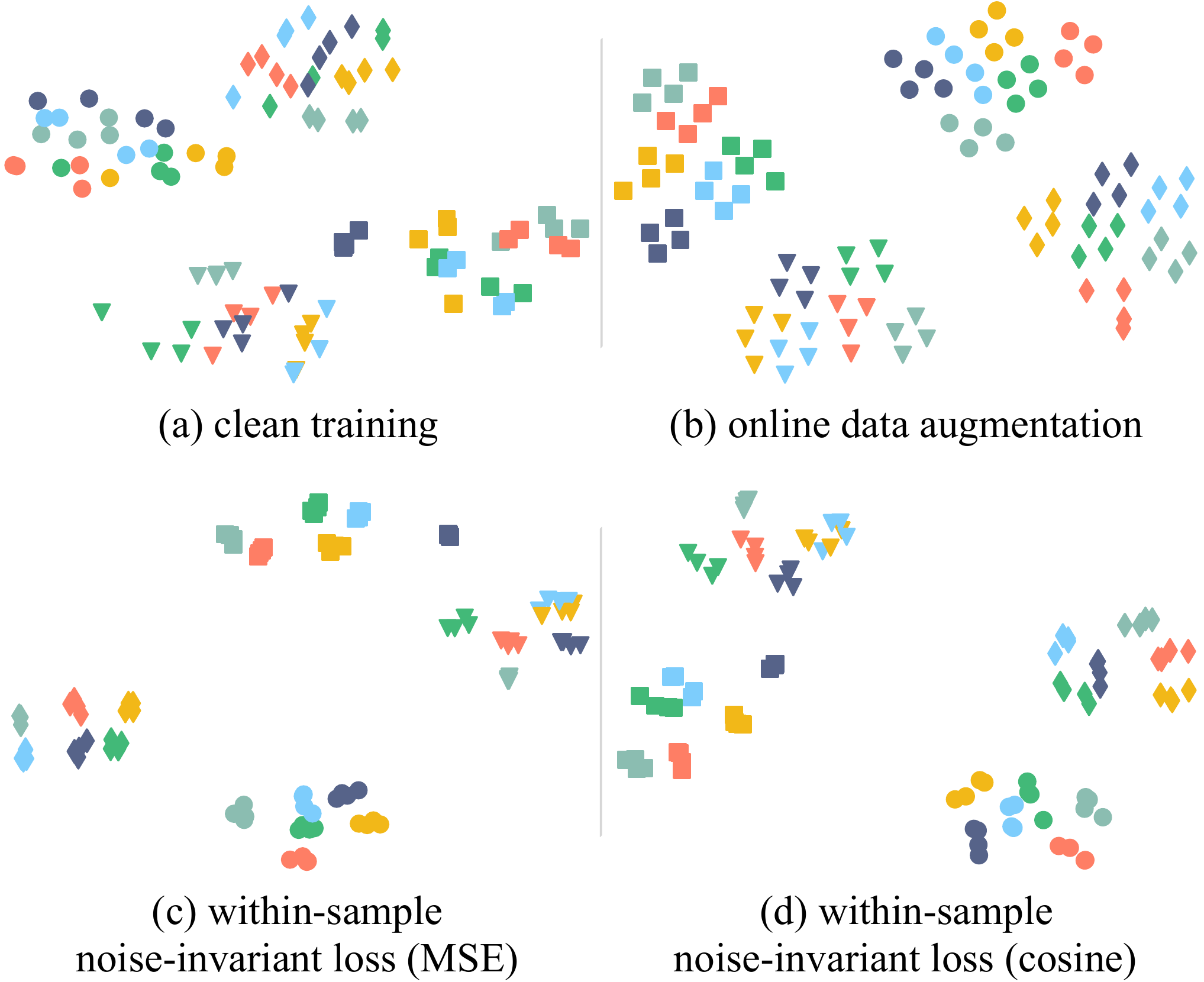}
  \caption{t-SNE visualization of speaker embeddings extracted from the training dataset. Each marker corresponds to a different speaker, and each color in the same marker corresponds to a different utterance. The clean utterances and its noisy copies have the same color.}
  \label{fig:tsne}
\end{figure}

Several observations from the results are discussed in the following. 1) The experimental results confirm that data augmentation strategy can greatly improve the performance of the deep speaker embedding system under noisy conditions. 2) Comparing with the offline data augmentation strategy, the performance improvement achieved by online data augmentation is more obvious in the low SNR conditions. 3) Training the deep speaker embedding system with within-sample variability-invariant loss can improve the system performance in the clean and all noisy conditions. 4) Comparing with the network trained with offline data augmentation, the proposed framework using within-sample variability-invariant loss with online data augmentation achieves 13.0\% and 6.5\% reduction in terms of EER and DCF respectively. 5) When the speaker embedding network is trained discriminatively using the A-softmax loss with angular margin, the proposed within-class loss can still improve the system performance by setting constraints on the distance among the clean utterance and its noisy copies.

The detection error tradeoff (DET) curves in figure \ref{fig:det_curve} provide comparisons among four selected systems, two of which are trained with our proposed framework. The DET curve uses testing trials from all the noisy conditions.

We also visualized the speaker embeddings by using the t-distributed stochastic neighbor embedding (t-SNE) algorithm \cite{maaten_visualizing_2008}. The two-dimensional results of the speaker embeddings are shown in figure \ref{fig:tsne}. Four speakers, each with six clean utterances, are selected from the training dataset for visualization. Also, each clean utterance has three 5dB noisy copies of music, babble and ambient noises. Comparing with the clean training condition, data augmentation helps the clean and noisy embeddings from the same utterance cluster together. Further, after training the deep speaker embedding network with within-noise variability-invariant loss, the clean and noisy embeddings of the same utterance are closer to each other. 

The loss values of each training epoch are shown in figure \ref{fig:loss} for the network with speaker softmax and within-sample MSE losses. The referenced MSE loss between embeddings from the clean and noisy data of the converged network trained with only softmax loss is also given. We can observe that the MSE loss is maintained at a low level during training, which helps the network to extract noisy embedding similar to its clean version.

\section{Conclusion}
This paper has proposed the within-sample variability-invariant loss for deep speaker embedding networks under noisy conditions. By setting constraints on the embeddings extracted from the clean utterance and its noisy copies, the proposed loss works with the original speaker identification loss to learn robust embedding for noisy speeches. We also employ the data preparation strategy of generating the clean and noisy utterance pairs on-the-fly to help the speaker embedding network generalize better under noisy environments. The proposed framework is flexible and can be extended to other similar applications when multiple views of the same training speech sample are available.

\section{Acknowledgement}
This research is funded in part by the National Natural Science Foundation of China (61773413) and Duke Kunshan University.

\bibliographystyle{IEEEbib}
\bibliography{mybib}

\end{document}